\documentclass[notitlepage,11pt]{article}%
\usepackage{amsfonts}
\usepackage{amsmath}%
\usepackage{amssymb}%
\usepackage{graphicx}
\newtheorem{theorem}{Theorem}

\newtheorem{corollary}[theorem]{Corollary}

\newtheorem{example}[theorem]{Example}

\newtheorem{lemma}[theorem]{Lemma}

\newtheorem{proposition}[theorem]{Proposition}
\newtheorem{remark}[theorem]{Remark}

\newenvironment{proof}[1][Proof]{\noindent\textbf{#1.} }{\ \rule{0.5em}{0.5em}}
\begin{document}

\title{ON\ THE\ NOTION\ OF\ PHASE IN\ \ MECHANICS}
\author{Maurice A de Gosson\\BTH-Karlskrona\\SE-371 79 Karlskrona}
\maketitle

\begin{abstract}
The notion of phase plays an essential role in both semiclassical and quantum
mechanics. But what is exactly a phase, and how does it change with time? It
turns out that the most universal definition of a phase can be given in terms
of Lagrangian manifolds by exploiting the properties of the
Poincar\'{e}--Cartan form. Such a phase is defined, not in configuration
space, but rather in phase space and is thus insensitive to the appearance of
caustics. Surprisingly enough this approach allows us to recover the
Heisenberg--Weyl formalism without invoking commutation relations for observables.

\end{abstract}
\tableofcontents

\section{Introduction}

What is a phase? A common conception is that it is something like an angle;
but this does of course not tell us very much concretely. Let us look up the
word \textquotedblleft phase\textquotedblright\ in the
Webster\footnote{Webster New Encyclopedia, 1994 edition.}. We find there that
\textquotedblright...\textit{[a phase is] the stage of progress in a regularly
recurring motion or a cyclic progress (as a wave or vibration) in relation to
a reference point.}\textquotedblright\ The last few words really go straight
to the point: the vocation of a phase is to describe a \textit{variation} --it
has no absolute meaning by itself. So what would then a good definition of the
variation of \textquotedblleft phase\textquotedblright\ be for a mechanical
system? Consider a Hamiltonian system (in $n$ degrees of freedom) with
Hamiltonian $H=H(x,p,t)$; here $x=(x_{1},...,x_{n})$, $p=(p_{1},...,p_{n})$.
We will define the variation of the phase of that system when it evolves from
a state $z^{\prime}=(x^{\prime},p^{\prime})$ at time $t^{\prime}$ to a state
$z=(x,p)$ at time $t$ by the formula%
\begin{equation}
\Delta\Phi=\int_{z^{\prime},t^{\prime}}^{z,t}p\mathrm{d}x-H\mathrm{d}%
t\label{delta}%
\end{equation}
where the integration is performed along the arc joining $(z^{\prime
},t^{\prime})$ to $(z,t)$ in time-dependent phase space, and determined by the
Hamilton equations for $H$.

So far, so good. But again: \textit{what is} then \textquotedblleft
the\textquotedblright\ phase of that system? A clue is given by
Hamilton--Jacobi's equation with initial datum%
\begin{equation}
\frac{\partial\Phi}{\partial t}+H(x,\nabla_{x}\Phi)=0\text{ \ , \ }%
\Phi(x,t^{\prime})=\Phi^{\prime}(x)\text{.}\label{HJ}%
\end{equation}
Assume that $H$ is of the classical type \textquotedblleft kinetic energy $+$
smooth potential\textquotedblright; then the solution of the problem
(\ref{HJ}) always exists (and is unique) if $|t-t^{\prime}|$ is sufficiently
small. This solution $\Phi=\Phi(x,t)$ is obtained as follows. Let us denote by
$(f_{t,t^{\prime}}^{H})$ the time-dependent flow determined by $H$ and
consider the graph $\mathbb{V}^{\prime}$ of the function $p^{\prime}%
=\nabla_{x}\Phi^{\prime}(x^{\prime})$. For small values of $|t-t^{\prime}|$
the image $\mathbb{V}=f_{t,t^{\prime}}^{H}(\mathbb{V}^{\prime})$ will still
project diffeomorphically on configuration space and hence still be a graph;
the coordinate $x$ being given let $p$ be the unique momentum vector such that
$z=(x,p)\in\mathbb{V}$ and define $z^{\prime}=(x^{\prime},p^{\prime}%
)\in\mathbb{V}^{\prime}$ by $z=f_{t,t^{\prime}}^{H}(z^{\prime})$; then the
difference $\Phi(x,t)-\Phi^{\prime}(x^{\prime})$ is the quantity%
\[
\Delta\Phi(x,x^{\prime})=\int_{z^{\prime},t^{\prime}}^{z,t}p\mathrm{d}%
x-H\mathrm{d}t\text{.}%
\]
and we can take the formula%
\begin{equation}
\Phi(x,t)=\Phi^{\prime}(x^{\prime})+\int_{z^{\prime},t^{\prime}}%
^{z,t}p\mathrm{d}x-H\mathrm{d}t\label{solution}%
\end{equation}
as a $\emph{definition}$ of the phase of the Hamiltonian system. Such a choice
is quite correct, and very much in the spirit of Hamilton--Jacobi theory. It
is however too restrictive,because the definition of $\Phi(x,t)$ heavily
relies on the fat that we were able to define a point $x^{\prime}$ on the
initial graph $\mathbb{V}^{\prime}$ via the formula $(x,p)=f_{t,t^{\prime}%
}^{H}(x^{\prime},p^{\prime})$. This is only possible if $\mathbb{V}%
=f_{t,t^{\prime}}^{H}(\mathbb{V}^{\prime})$ is itself is a graph, and this is
in general no longer the case when $|t-t^{\prime}|$ becomes too large: for a
given $x$ there will perhaps be several points $(x,p_{1}),(x,p_{2}),...,$ of
$\mathbb{V}$ having the same position coordinate due to the \textquotedblright
bending\textquotedblright\ of $\mathbb{V}^{\prime}$ by the flow as time
elapses, and formula (\ref{solution}) will no longer make sense (to use an
older terminology, the phase becomes \textquotedblleft
multi-valued\textquotedblright). This is the usual problem to which one is
confronted to in Hamilton--Jacobi theory, and is also, by the way, one of the
reasons for which the \textit{WKB}\ method breaks down for large times: the
semiclassical solutions to Schr\"{o}dinger's equation one wants to define on
the set $\mathbb{V}=f_{t,t^{\prime}}^{H}(\mathbb{V}^{\prime})$ blow up because
of the appearance of \textquotedblleft caustics\textquotedblright\ related to
that bending. In semiclassical mechanics, the remedy to this situation is
well-known: one renounces to the usual solutions of Hamilton--Jacobi's
equation (\ref{HJ}) and one considers the manifolds $\mathbb{V}$ themselves
--whether they are graphs, or not-- as generalized solutions: this is the
phase space approach to semiclassical mechanics inaugurated by Keller, and
further developed by Maslov \cite{Maslov}, Maslov and Fedoriuk \cite{MF},
Leray \cite{Leray}, and many others. Now, these manifolds are not arbitrary;
they are \emph{Lagrangian submanifolds} of phase space. These special
manifolds can be thought as generalizations of the usual invariant tori of
Liouville integrable systems of Hamiltonian mechanics, but there use is
certainly not limited to this venerable topic: Lagrangian manifolds have a
life of their own, and intervene in various fields. Even if one doesn't have
to take Weinstein's \cite{wein1} creed \textquotedblleft\textit{everything is
a Lagrangian manifold!}\textquotedblright\ quite at face value, it is however
true that Lagrangian manifolds can be associated in a very natural way both to
classical and quantum systems. (We will discuss this in some detail in Section
\ref{de}; in any case the solution of Cauchy's problem for Hamilton--Jacobi's
equation anyway involves \textit{de facto} a Lagrangian manifold, whether the
system is Liouville integrable or not.) The situation is even more clear-cut
in quantum mechanics: to every quantum system whose evolution is governed by
Schr\"{o}dinger's equation%
\[
i\hbar\frac{\partial\Psi}{\partial t}=\hat{H}\Psi
\]
one can associate a canonical Lagrangian manifold: writing the wavefunction in
polar form $\Psi=\exp(i\Phi/\hbar)$ the graph $p=\nabla_{x}\Phi(x,t) $ of the
phase at time $t$ is a Lagrangian manifold. (We have used this fact in
\cite{IOP} to show how this can be used to understand Schr\"{o}dinger's
equation in the framework of the Hamilton--Jacobi formalism.)

An essential tool for the study of the time-evolution of the phase of a
Lagrangian manifold under the action of Hamiltonian flows is the
Poincar\'{e}--Cartan form $p\mathrm{d}x-H\mathrm{d}t$. Its importance comes
from the fact that it is a (relative) integral invariant. Strangely enough,
this property is often mentioned in both the mathematical and physical
literature, but seldom fully exploited. Admittedly, the approach
\textquotedblleft Lagrangian manifolds $+$ Poincar\'{e}--Cartan
invariant\textquotedblright\ is certainly not new; for instance Weinstein
\cite{wein2} has used it to study the global properties of paths of Lagrangian
manifold subject to an \textquotedblleft isodrastic\textquotedblright\ (that
is, action-preserving) deformation; due to the heavy use of intrinsic
differential geometry Weinstein's paper is however not easily accessible to a
physical audience. On the other hand many of the results contained in Section
\ref{de} can be found in an elusive or fragmentary form elsewhere
(\textit{e.g.} \cite{Wiley,Leray}). In Section \ref{qua} we show how the
properties of the phase allows us to recover the Heisenberg--Weyl operator
formalism familiar from semiclassical mechanics.

This article is relatively self-contained: the proofs are complete (even if
concise), and we have found it useful to shortly review the necessary topics
from symplectic geometry and Hamiltonian mechanics (the invariance property of
the Poincar\'{e}--Cartan form is one example); we refer the reader to the
classical treatises \cite{Arnold,Tal,AM,Marle} (cited in increasing order of
mathematical difficulty) for the notions of differential geometry that we will
use.\bigskip

\noindent\textbf{Notations.} The phase space $\mathbb{R}_{z}^{2n}%
=\mathbb{R}_{x}^{n}\times\mathbb{R}_{p}^{n}$ is equipped with the standard
symplectic form $\sigma$:
\[
\sigma(z,z^{\prime})=px^{\prime}-p^{\prime}x=\sum_{j=1}^{n}p_{j}x_{j}^{\prime
}-p_{j}^{\prime}x_{j}%
\]
if $z=(x,p)$, $z^{\prime}=(x^{\prime},p^{\prime})$; in differential notation:%
\[
\sigma=\mathrm{d}p\wedge\mathrm{d}x=\sum_{j=1}^{n}\mathrm{d}p_{j}%
\wedge\mathrm{d}x_{j}\text{.}%
\]
\ \ 

A Lagrangian plane is a $n$-dimensional linear subspace $\ell$ of
$\mathbb{R}_{z}^{2n}$ such that the symplectic form $\sigma$ vanishes on every
pair of vectors of $\ell$:%
\[
z,z^{\prime}\in\ell\Longrightarrow\sigma(z,z^{\prime})=0\text{.}%
\]
Equivalent definitions are: $(\ast)$ a Lagrangian plane is the image of
configuration space $\mathbb{R}_{x}^{n}$ (or momentum space $\mathbb{R}%
_{p}^{n}$) by a\ linear symplectic transformation (\textit{i.e.} a symplectic
matrix); $(\ast\ast)$ A $n$-plane with equation $Ax+Bp=0$ is Lagrangian if and
only if $A^{T}B=BA^{T}$.

In all what follows the letter $\mathbb{V}$ will denote a connected (but not
necessarily compact) Lagrangian submanifold of the phase space $\mathbb{R}%
_{z}^{2n}$, that is:

\begin{itemize}
\item $\mathbb{V}$ has dimension $n$ as a manifold;

\item the tangent space $\ell(z)=T_{z}\mathbb{V}$ at every point $z$ of
$\mathbb{V}$ is a Lagrangian plane.
\end{itemize}

\section{Lagrangian Manifolds in Mechanics\label{de}}

A basic (but not generic) example of Lagrangian manifold is the following: let
$\Phi=\Phi(x)$ be a smooth function defined on some open domain in
configuration space. Then
\[
\mathbb{V}:p=\nabla_{x}\Phi(x)
\]
is a Lagrangian manifold (sometimes called \textquotedblleft exact Lagrangian
manifold\textquotedblright). The image of a Lagrangian manifold $\mathbb{V}$
by a symplectic diffeomorphism $f$ is again a Lagrangian manifold:
$f(\mathbb{V)}$ is a manifold, and the tangent mapping $\mathrm{d}f(z_{0})$ is
an isomorphism of $\ell(z_{0})=T_{z_{0}}\mathbb{V}$ on $\ell(f(z_{0}%
))=T_{f(z_{0})}f(\mathbb{V)}$; that isomorphism is symplectic hence
$\ell(f(z_{0})$ is a Lagrangian plane. Observe that a Lagrangian plane is a
Lagrangian manifold in it own right, and so is the image by a Lagrangian plane
by a symplectic diffeomorphism.

Let us begin by making the following pedestrian --but important-- remark.
Suppose that we have a system of $N$ point-like particles at some time, say
$t=0$, and that we know all the positions and momenta of these particles; that
system is thus identified with a point $z=(x,p)$ in phase space. We can always
find a Lagrangian manifold (in fact, infinitely many) carrying this point $z$.
The easiest example is obtained by choosing numbers $a_{1},...,a_{n}$ such
that $p_{j}=a_{j}x_{j}$; denoting by $M$ the diagonal matrix with diagonal
entries $a_{j}$, the linear space $\ell:p=Mx$ is a Lagrangian plane (and
hence, a fortiori, a Lagrangian manifold).

We can even do better: assume that the system of $N$ particles is Hamiltonian
with Hamiltonian $H$, and let $E$ be the energy of the system. Consider a
solution $\Phi=\Phi(x)$ of the reduced Hamilton--Jacobi equation%
\[
H(x,\nabla_{x}\Phi)=E\text{.}%
\]
The manifold $\mathbb{V}:p=\nabla_{x}\Phi(x)$ is Lagrangian and the energy of
$H$ is constant on it; it thus lies on the energy shell $\Sigma_{E}:H(z)=E$.
We thus see that independently of any integrability condition one can
associate an exact Lagrangian manifold to every Hamiltonian system; that
Lagrangian manifold can be interpreted as a set carrying a \textquotedblleft
cloud\textquotedblright\ of particles, in fact a statistical ensemble where
the positions and momenta are in interrelated by the formula $p=\nabla_{x}%
\Phi(x)$. More generally, there is no need to assume ha there is such a
correlation, and one can as well consider a Lagrangian manifold as a set
representing a physical state. When one weights this manifold by a density (or
rather a de Rham form, see \cite{ICP,paselect}) and thereafter imposing to it
the Maslov(or \textit{EBK}) quantum conditions, one obtains semiclassical mechanics.

It should be noted that Lagrangian (sub) manifolds actually play an ubiquitous
role in physics. For instance, the role of \textquotedblleft reciprocity
laws\textquotedblright\ giving arise to such manifolds in thermodynamics
(\textquotedblleft Onsager relations\textquotedblright), thermostatics
(\textquotedblleft Maxwell relations\textquotedblright), and in electricity
and electromagnetism is well-known. Tulczyjew and Oster actually view
Lagrangian manifolds as the basic entities describing physical systems (see
Abraham and Marsden \cite{AM}, Ch. 5, for an extensive list of references and
many examples).

\section{The Phase of a Lagrangian Submanifold}

et $\mathbb{V}$ be an exact Lagrangian manifold, defined by the equation
$p=\nabla_{x}\Phi(x)$. Then $\varphi(z)=\Phi(x)$ is a phase of $\mathbb{V} $.
This is obvious since%
\[
\mathrm{d}\varphi(z)=\mathrm{d}\Phi(x)=p\mathrm{d}x\text{.}%
\]
We observe that the phase can be expressed as an integral:%
\[
\varphi(z)=\Phi(x_{0})+\int_{\gamma}\mathrm{d}\Phi(x)
\]
where $\gamma$ is any path in configuration space joining $x_{0}$ to $x$.

To see what a notion of phase could be for a Lagrangian manifold which is not
a graph, let us begin with a simple example. We would like to define on the
circle $S^{1}(R):x^{2}+p^{2}=R^{2}$ in the plane $\mathbb{R}_{z}^{2}$ a smooth
function $\varphi$ whose differential $\mathrm{d}\varphi$ is the action form
$p\mathrm{d}x$. Passing to polar coordinates $x=R\cos\theta$, $p=R\sin\theta$
the condition $\mathrm{d}\varphi=p\mathrm{d}x$ is%
\[
\mathrm{d}\varphi(\theta)=-R^{2}\sin^{2}\theta
\]
which, integrated, leads to
\begin{equation}
\varphi(\theta)=\frac{R^{2}}{2}(\cos\theta\sin\theta-\theta)\text{.}%
\label{circle}%
\end{equation}
Now, that function is not defined on the circle itself, because $\varphi
(\theta+2\pi)=\varphi(\theta)-\pi R^{2}\neq\varphi(\theta)$. We can however
view $\varphi(\theta)$ as defined on the universal covering of $S^{1}(R)$,
identified with the real line $\mathbb{R}_{\theta}$, the projection
$\pi:\mathbb{R}_{\theta}\longrightarrow S^{1}(R)$ being given by $\pi
(\theta)=(R\cos\theta,R\sin\theta)$.

Consider, more generally, a completely integrable system with Hamiltonian $H$,
and $(\theta,I)=(\theta_{1},...,\theta_{n};I_{1},...,I_{n})$ the corresponding
angle-action variables. We have $H(x,p)=K(I)$ and the motion is given by
\[
\theta(t)=\theta(0)+\omega(I(0))t\text{ \ \ , \ \ }I(t)=I(0)
\]
where the frequency vector $\omega(I)=(\omega_{1}(I),...,\omega_{n}(I))$ is
the gradient of $K$: $\omega(I)=\nabla_{I}K(I)$. The motion takes place on the
Lagrangian manifold $I(t)=I(0)$. Topologically this manifold is identified
with a product of $n$ unit circles, each lying in a plane of conjugate
variables. Recalling that $\theta=(\theta_{1},...,\theta_{n})$ the phase of
$\mathbb{T}$ is thus%
\[
\varphi(\theta)=\frac{1}{2}\sum_{j=1}^{n}(\cos\theta_{j}\sin\theta_{j}%
-\theta_{j})
\]
in view of (\ref{circle}).

Consider now an \emph{arbitrary} Lagrangian manifold $\mathbb{V}$, and choose
a \textquotedblleft base point\textquotedblright\ $\bar{z}=(\bar{x},\bar{p})$
on $\mathbb{V}$; we denote by $\pi_{1}(\mathbb{V})$ the fundamental group
$\pi_{1}(\mathbb{V},\bar{z})$. Let us denote by $\mathbb{\check{V}}$ the set
of all homotopy classes $\check{z}$ of paths $\gamma(\bar{z},z)$ starting at
$\bar{z}$ and ending at $z$, and by $\pi:\mathbb{\check{V}\longrightarrow V}$
the mapping which to $\check{z}$ associates the endpoint $z$ of any of its
representatives $\gamma(\bar{z},z)$. The set $\mathbb{\check{V}}$ can be
equipped with a topology having the following properties: $(\ast)$
$\mathbb{\check{V}}$ is simply connected; $(\ast\ast) $ $\pi$ is a covering
mapping: every $z\in\mathbb{V}$ has an open neighborhood $U$ such that
$\pi^{-1}(U)$ is the disjoint union of a sequence of open sets $\check{U}%
_{1},\check{U}_{2},...$ such that the restriction of $\pi$ to each of the
$\check{U}_{j}$ is a diffeomorphism onto $U$. With that topology and
projection, $\mathbb{\check{V}}$ is the universal covering of $\mathbb{V}$.
Consider now the action form
\[
p\mathrm{d}x=p_{1}\mathrm{d}x_{1}+\cdot\cdot\cdot+p_{n}\mathrm{d}x_{n}%
\]
on $\mathbb{V}$; we can \textquotedblleft pull-back\textquotedblright\ this
form to $\mathbb{\check{V}}$ using the projection $\pi$, thus obtaining a
one-form $\pi^{\ast}(p\mathrm{d}x)$. Now%
\[
\mathrm{d}\pi^{\ast}(p\mathrm{d}x)=\pi^{\ast}\mathrm{d}(p\mathrm{d}%
x)=\pi^{\ast}(\mathrm{d}p\wedge\mathrm{d}x)
\]
and $\mathrm{d}p\wedge\mathrm{d}x=\sigma$ is identically zero on $\mathbb{V}$,
hence the form $\pi^{\ast}(p\mathrm{d}x)$ is closed on $\mathbb{\check{V}}$.
Since $\mathbb{\check{V}}$ is contractible $\pi^{\ast}(p\mathrm{d}x)$ is an
exact form on $\mathbb{\check{V}}$ in view of Poincar\'{e}'s lemma and we can
thus find infinitely many functions $\varphi:\mathbb{\check{V}\longrightarrow
R}$, all differing by a constant, such that $\mathrm{d}\varphi(\check{z}%
)=\pi^{\ast}(p\mathrm{d}x)$. Making a slight abuse of notation by identifying
$p\mathrm{d}x$ and its pull-back $\pi^{\ast}(p\mathrm{d}x)$ we can summarize
the discussion above as follows:

\begin{quotation}
\textit{There exists a differentiable function }$\varphi:\check{V}%
\longrightarrow R$\textit{\ such that}
\begin{equation}
\mathrm{d}\varphi(\check{z})=p\mathrm{d}x\text{ \ \ \textrm{if} \ \ }%
\pi(\check{z})=z=(x,p)\text{.}\label{ph3}%
\end{equation}

\end{quotation}

We will call such a function \textit{a }\emph{phase}\textit{\ of }$\mathbb{V}
$, although $\varphi$ is in general defined on the universal covering
$\mathbb{\check{V}}$. Notice that we can always fix one such phase by imposing
a given value at some point of $\mathbb{\check{V}}$; for instance we can
choose $\varphi(\bar{z})=0$ where $\bar{z}$ is identified with the (homotopy
class of) the constant loop $\gamma(\bar{z},z)$.

A straightforward example of phase one can associate to a system of particles
represented by a phase space point $z$ is the following:

\begin{example}
Let $\ell:p=Mx$ be a Lagrangian plane passing through $z$ (that such a plane
always exists was discussed above). Choosing the origin as base point:
$\bar{z}=0$ the phase is
\[
\phi(z)=\tfrac{1}{2}p\cdot x=\tfrac{1}{2}Mx^{2}%
\]
(where we have set $Mx^{2}=Mx\cdot x$).
\end{example}

Phases on Lagrangian manifolds can be explicitly constructed by integrating
the action form along paths:

\begin{proposition}
Let $z$ be any point of $\mathbb{V}$ and $\gamma(\bar{z},z)$ an arbitrary
continuous path in $\mathbb{V}$ joining $\bar{z}$ to $z$. The line integral%
\[
I(z)=\int_{\gamma(\bar{z},z)}p\mathrm{d}x
\]
only depends on the homotopy class $\check{z}$ of $\gamma(\bar{z},z)$ and
defines a phase of $\mathbb{V}$.
\end{proposition}

\begin{proof}
Let $\gamma^{\prime}(\bar{z},z)$ be another path joining $\bar{z}$ to $z$ in
$\mathbb{V}$ and homotopic to $\gamma(\bar{z},z)$; the loop $\delta
=\gamma(\bar{z},z)-\gamma^{\prime}(\bar{z},z)$ is thus homotopic to a point in
$\mathbb{V}$. Let $h=h(s,t)$, $0\leq s,t\leq1$ be such a homotopy:
$h(0,t)=\delta(t)$, $h(1,t)=0$. As $s$ varies from $0$ to $1$ the loop
$\delta$ will sweep out a two-dimensional surface $\mathcal{D}$ with boundary
$\delta$ contained in $\mathbb{V}$. In view of\ the multi-dimensional Stokes
theorem we have%
\[
\int_{\delta}p\mathrm{d}x=\int\int_{\mathcal{D}}\mathrm{d}p\wedge\mathrm{d}x=0
\]
where the last equality follows from the fact that $\mathcal{D}$ is a subset
of a Lagrangian manifold. It follows from this equality that
\[
\int_{\gamma(\bar{z},z)}p\mathrm{d}x=\int_{\gamma^{\prime}(\bar{z}%
,z)}p\mathrm{d}x
\]
hence the integral of $p\mathrm{d}x$ along $\gamma(\bar{z},z)$ only depends on
the homotopy class in $\mathbb{V}$ of the path joining $\bar{z}$ to $z$; it is
thus a function of $\check{z}\in\mathbb{\check{V}}$. There remains to show
that the function $\varphi:\mathbb{\check{V}\longrightarrow R}$ defined by%
\begin{equation}
\varphi(\check{z})=\int_{\gamma(\bar{z},z)}p\mathrm{d}x\label{ph1}%
\end{equation}
is such that $\mathrm{d}\varphi(\check{z})=p\mathrm{d}x$. The property being
local, we can assume that $\mathbb{V}$ is simply connected, so that
$\mathbb{\check{V}=V}$. Since $\mathbb{V}$ is diffeomorphic to $\ell
(z)=T_{z}\mathbb{V}$ in a neighborhood of $z$ we can reduce the proof to the
case where $\mathbb{V}$ is a Lagrangian plane $\ell$. Let $Ax+Bp=0$
($A^{T}B=BA^{T}$) be an equation of $\ell$, and
\[
\gamma(z):t\longmapsto(-B^{T}u(t),A^{T}u(t))\text{ \ },\text{ \ }0\leq t\leq1
\]
be a differentiable curve starting from $0$ and ending at $z=(-B^{T}%
u(1),A^{T}u(1))$. We have%
\begin{align*}
\varphi(z)  & =\int_{\gamma(z)}p\mathrm{d}x\\
& =-\int_{0}^{1}A^{T}u(t)\cdot B^{T}\dot{u}(t)\mathrm{d}t\\
& =-\int_{0}^{1}BA^{T}u(t)\cdot\dot{u}(t)\mathrm{d}t
\end{align*}
and hence, since $BA^{T}$ is symmetric:%
\[
\varphi(z)=-\tfrac{1}{2}BA^{T}u(1)^{2}%
\]
that is%
\[
\mathrm{d}\varphi(z)=-BA^{T}u(1)\mathrm{d}u(1)=p\mathrm{d}x\text{.}%
\]

\end{proof}

As already observed above we are slightly abusing language by calling
$\varphi$ a \textquotedblleft phase of $\mathbb{V}$\textquotedblright\ since
$\varphi$ is multi-valued on $\mathbb{V}$. This multi-valuedness is made
explicit by studying the action of $\pi_{1}(\mathbb{V})$ on $\mathbb{\check
{V}}$. The latter is defined as follows: let $\gamma$ be a loop in
$\mathbb{V}$ with origin $z_{0}$ and $\check{\gamma}\in\pi_{1}(\mathbb{V})$
its homotopy class. Then $\check{\gamma}\check{z}$ is the homotopy class of
the loop $\gamma$ followed by the path $\gamma(z)$ representing $\check{z}$.
From the definition of the phase $\varphi$ follows that
\begin{equation}
\varphi(\check{\gamma}\check{z})=\varphi(\check{z})+\oint_{\gamma}%
p\mathrm{d}x\text{.}\label{ph2}%
\end{equation}
The phase is thus defined on $\mathbb{V}$ itself if and only if $\int_{\gamma
}p\mathrm{d}x=0$ for all loops in $\mathbb{V}$; this is the case if
$\mathbb{V}$ is contractible. However Gromov has proved in \cite{Gromov} that
if $\mathbb{V}$ is closed (\textit{i.e.} compact and without boundary) then we
cannot have $\oint_{\gamma}p\mathrm{d}x=0$ for all loops $\gamma$ in
$\mathbb{V}$; to construct the phase of such a manifold we thus have to use
the procedure above.

\section{The Local Expression of the Phase}

Recall that a Lagrangian manifold which can be represented by an equation
$p=\nabla_{x}\Phi(x)$ is called an \textquotedblleft exact Lagrangian
manifold\textquotedblright. It turns out that Lagrangian manifolds are
(locally) exact outside their caustic set, and this is most easily described
in terms of the phase defined above. We use the following standard
terminology: a point $z$ of a Lagrangian manifold $\mathbb{V}$ is called a
\textquotedblleft caustic point\textquotedblright\ if $z$ has no neighbourhood
in $\mathbb{V}$ for which the restriction of the mapping $z=(x,p)\longmapsto
x$ is a diffeomorphism; at a caustic point the tangent space $\ell
(z)=T_{x}\mathbb{V}$ is the momentum space $0\times\mathbb{R}_{p}^{n}$. The
set $\Sigma$ of all caustic points of $\mathbb{V}$ is called the \emph{caustic
of} $\mathbb{V}$. Of course, caustics have no intrinsic meaning, whatsoever:
there are just artefacts coming from the choice of a privileged $n$%
-dimensional plane (\textit{e.g.}, the configuration space) on which one
projects the motion.

Let $\mathbb{U}$ be an open subset of $\mathbb{V}$ which contains no caustic
points: $\mathbb{U}\cap\Sigma=\emptyset$. Then the restriction $\chi
_{\mathbb{U}}$ to $\mathbb{U}$ of the projection $\chi:(x,p)\longmapsto x$ is
a diffeomorphism of $\mathbb{U}$ onto its image $\chi_{\mathbb{U}}%
(\mathbb{U})$, and $(\mathbb{U},\chi_{\mathbb{U}})$ is thus a local chart of
$\mathbb{V}$. Choosing $\mathbb{U}$ small enough, we can assume that the fibre
$\pi^{-1}(\mathbb{U})$ is the disjoint union of a family of open sets
$\mathbb{\check{U}}$ in the universal covering of $\mathbb{V}$ and such that
the restriction $\pi_{\mathbb{U}}$ to $\mathbb{\check{U}}$ of the projection
$\pi:\mathbb{\check{V}}\longrightarrow\mathbb{V}$ is a diffeomorphism onto
$\mathbb{U}$. It follows that $(\mathbb{\check{U}},\chi_{\mathbb{U}}\circ
\pi_{\mathbb{U}})$ is a local chart of $\mathbb{\check{V}}$.

\begin{proposition}
\label{fi}Let $\Phi$ be the local expression of the phase $\varphi$ in any of
the local charts $(\mathbb{\check{U}},\chi_{\mathbb{U}}\circ\pi_{\mathbb{U}}%
)$:%
\begin{equation}
\Phi(x)=\varphi((\chi_{\mathbb{U}}\circ\pi_{\mathbb{U}})^{-1}(x))\text{.}%
\label{fix}%
\end{equation}
The Lagrangian submanifold $\mathbb{U}$ is exact and can be represented by the
equation
\begin{equation}
p=\nabla_{x}\Phi(x)=\nabla_{x}\varphi((\chi_{\mathbb{U}}\circ\pi_{\mathbb{U}%
})^{-1}(x))\text{.}\label{pix}%
\end{equation}

\end{proposition}

\begin{proof}
Let us first show that the equation (\ref{pix}) remains unchanged if we
replace $(\mathbb{\check{U}},\chi_{\mathbb{U}}\circ\pi_{\mathbb{U}})$ by a
chart $(\mathbb{\check{U}}^{\prime},\chi_{\mathbb{U}^{\prime}}\circ
\pi_{\mathbb{U}^{\prime}})$ such that $\pi(\mathbb{\check{U}}^{\prime}%
)=\pi(\mathbb{\check{U}})$. There exists $\gamma\in\pi_{1}(\mathbb{V})$ such
that $\mathbb{\check{U}}^{\prime}=\gamma\mathbb{\check{U}}$ hence, by
(\ref{ph2}), the restrictions $\varphi_{\mathbb{\check{U}}^{\prime}}$ and
$\varphi_{\mathbb{\check{U}}}$ differ by the constant
\[
C(\gamma)=\oint_{\gamma}p\mathrm{d}x\text{.}%
\]
It follows that%
\[
\nabla_{x}\varphi((\chi_{\mathbb{U}^{\prime}}\circ\pi_{\mathbb{U}^{\prime}%
})^{-1}(x))=\nabla_{x}\varphi((\chi_{\mathbb{U}}\circ\pi_{\mathbb{U}}%
)^{-1}(x))
\]
and hence the right-hand side of the identity (\ref{pix}) does not depend on
the choice of local chart $(\mathbb{\check{U}},\chi_{\mathbb{U}}\circ
\pi_{\mathbb{U}})$. Set now $(\chi_{\mathbb{U}}\circ\pi_{\mathbb{U}}%
)^{-1}(x)=(p(x),x)$; we have, for $x\in\chi_{\mathbb{U}}\circ\pi_{\mathbb{U}%
}(\mathbb{U)}$,%
\[
\mathrm{d}\Phi(x)=\mathrm{d}\varphi(p(x),x)=p(x)\mathrm{d}x
\]
hence (\ref{pix}).
\end{proof}

\section{Symplectic Frames and Lagrangian Phases\label{frame}}

The observant reader will have noticed that the phase of a Lagrangian manifold
was defined in terms of one special coordinate system, namely the canonical
coordinates $x,p$. It is of course of interest to determine what happens to
the phase under symplectic changes of variables. Let us introduce, following
Leray \cite{Leray}, the notion of symplectic frame: by definition, a
symplectic frame is any pair $(\ell,\ell^{\ast})$ of Lagrangian planes such
that $\mathbb{R}_{z}^{2n}=\ell\oplus\ell^{\ast}$; equivalently: $\ell\cap
\ell^{\ast}=0$. Set $\ell_{x}=\mathbb{R}_{x}^{n}\times0$ and $\ell_{p}%
=0\times\mathbb{R}_{p}^{n}$ (the configuration space, and the momentum space,
respectively). The pair $(\ell_{x},\ell_{p})$ is a symplectic frame: we call
it the canonical frame. The symplectic group acts transitively on all pairs of
transverse Lagrangian planes (see \cite{Wiley,ICP}); it follows that the image
$S(\ell,\ell^{\ast})=(S\ell,S\ell^{\ast})$ of a symplectic frame is a
symplectic frame, and that for every pair $(\ell,\ell^{\ast})$, $(\ell
^{\prime},\ell^{\prime\ast})$ of symplectic frame there exists $R\in
\operatorname*{Sp}(n)$ such that $(\ell,\ell^{\ast})=R(\ell^{\prime}%
,\ell^{\prime\ast}) $ (i.e. $\ell=R\ell^{\prime}$ and $\ell^{\ast}%
=R\ell^{\prime\ast}$). We will call such an $R$ a \textit{symplectic change of
frame}; a manifold which is Lagrangian in one such frame is Lagrangian in all
symplectic frames and we will see that there is an intrinsic (\textit{i.e.}
frame-independent) function on $\mathbb{\check{V}}$ which we call, again
following Leray, the \textit{Lagrangian phase} of $\mathbb{V}$.

For the sake of notational brevity we will omit the dot $\cdot$ for scalar
products and write, for instance, $px$ in place of $p\cdot x$.

Let $\operatorname*{Sp}(n)$ be the symplectic group: $S\in\operatorname*{Sp}%
(n)$ if and only if $S$ is a linear automorphism of $\mathbb{R}_{z}^{2n}$
preserving the symplectic form $\sigma$: $\sigma(Sz,Sz^{\prime})=\sigma
(z,z^{\prime})$ for all vectors $z,z^{\prime}$. For every $S\in
\operatorname*{Sp}(n)$ the image $S(\mathbb{V)}$ is also a Lagrangian
manifold. The following result allows us to compare the phases of $\mathbb{V}$
and $S(\mathbb{V)}$; it will also allow us to give a frame-independent
definition of the phase of $\mathbb{V}$.

\begin{proposition}
\label{un} For $S\in\operatorname*{Sp}(n)$ set $(x_{S},p_{S})=S(x,p)$.
$(\ast)$ We have
\begin{equation}
p_{S}\mathrm{d}x_{S}-x_{S}\mathrm{d}p_{S}=p\mathrm{d}x-x\mathrm{d}%
p\text{.}\label{ps}%
\end{equation}
$(\ast\ast)$ Define a function $\varphi_{S}:\mathbb{\check{V}}\longrightarrow
\mathbb{R}$ by the formula%
\begin{equation}
\varphi_{S}(\check{z})=\varphi(\check{z})+\tfrac{1}{2}(p_{S}x_{S}%
-px)\text{.}\label{ph4}%
\end{equation}
That function is differentiable, and we have
\begin{equation}
\mathrm{d}\varphi_{S}(\check{z})=p_{S}\mathrm{d}x_{S}\text{ \ \textrm{if}
\ }\pi(\check{z})=(x,p)\text{.}\label{defi}%
\end{equation}

\end{proposition}

\begin{proof}
$(\ast)$ Writing $S$ in block-matrix form%
\[
S=%
\begin{pmatrix}
A & B\\
C & D
\end{pmatrix}
\]
the condition that $S$ is symplectic implies that $A^{T}C$ and $B^{T}D$ are
symmetric, and that $A^{T}D-C^{T}B=I$. Setting $x_{S}=Ax+Bp$, $p_{S}=Cx+Dp$,
and expanding the products, we get%
\begin{align*}
p_{S}\mathrm{d}x_{S}-x_{S}\mathrm{d}p_{S} &  =(A^{T}Cx+A^{T}Dp-C^{T}%
Ax-C^{T}Bp)\mathrm{d}x+\\
&  (B^{T}Cx+B^{T}Dp-D^{T}Ax-D^{T}Bp)\mathrm{d}p\\
&  =p\mathrm{d}x-x\mathrm{d}p
\end{align*}
proving (\ref{ps}). (Notice that in general we do \textit{not have }%
$p_{S}\mathrm{d}x_{S}=p\mathrm{d}x$.). $(\ast\ast)$ Differentiating the
right-hand side of (\ref{ph4}) we get, since $\mathrm{d}\varphi(\check
{z})=p\mathrm{d}x$,
\begin{align*}
\mathrm{d}\varphi_{S}(\check{z})  & =\tfrac{1}{2}(p\mathrm{d}x-x\mathrm{d}%
p)+\tfrac{1}{2}\mathrm{d}(p_{S}x_{S})\\
& =\tfrac{1}{2}(p_{S}\mathrm{d}x_{S}-x_{S}\mathrm{d}p_{S})+\tfrac{1}%
{2}\mathrm{d}(p_{S}x_{S})\\
& =p_{S}\mathrm{d}x_{S}%
\end{align*}
which proves (\ref{defi}).
\end{proof}

We can identify the universal covering of $S(\mathbb{V)}$ with that,
$\mathbb{\check{V}}$, of $\mathbb{V}$: for this it suffices to define the
projection
\[
\pi_{S}=S\circ\pi:\mathbb{\check{V}\longrightarrow}S(\mathbb{V)}\text{ \ :
\ }\pi_{S}(\check{z})=Sz=(x_{S},p_{S})\text{.}%
\]
Proposition \ref{un} can then be restated as follows:

\begin{quotation}
\textit{The phase of }$S(V)$\textit{\ is the function }$\varphi_{S}:\check
{V}\longrightarrow R$\textit{\ defined by formula (\ref{ph4}): we have
}$d\varphi_{S}(\check{z})=pdx$\textit{\ if }$\pi_{S}(\check{z})=(x,p)$%
\textit{.}
\end{quotation}

We will call \textquotedblleft Lagrangian phase of $\mathbb{V}$%
\textquotedblright\ the function $\lambda:\mathbb{\check{V}}\longrightarrow
\mathbb{R}$ defined by%
\begin{equation}
\lambda(\check{z})=\varphi(\check{z})-\tfrac{1}{2}px\text{ \ \textrm{if}
\ }\pi(\check{z})=(x,p)\text{.}\label{laz1}%
\end{equation}
In view of Lemma \ref{un} the invariant phase $\lambda_{R}$ of the Lagrangian
manifold $R\mathbb{V}$ is
\[
\lambda_{R}(\check{z})=\varphi_{R}(\check{z})-\tfrac{1}{2}p_{R}x_{R}\text{
\ \textrm{if} \ }\pi_{R}(\check{z})=(x,p)\text{;}%
\]
since in view of formula (\ref{ph4}) we have%
\begin{equation}
\varphi_{R}(\check{z})=\varphi(\check{z})+\tfrac{1}{2}(p_{R}x_{R}%
-px)\label{laz3}%
\end{equation}
it follows that $\lambda_{R}(\check{z})=\lambda(\check{z})$: the Lagrangian
phase is thus the same in all symplectic frames.

Notice that it follows from definition (\ref{laz1}) that the differential of
the Lagrangian phase is%
\begin{equation}
\mathrm{d}\lambda(\check{z})=\tfrac{1}{2}(p\mathrm{d}x-x\mathrm{d}%
p)\text{.}\label{laz2}%
\end{equation}

Let us note the following particular case of Lemma \ref{un}: assume that $S$
is a free symplectic matrix, that is
\[
S=%
\begin{pmatrix}
A & B\\
C & D
\end{pmatrix}
\text{ \ \ , \ \ }\det B\neq0
\]
(equivalently, $S(0\times\mathbb{R}_{p}^{n})\cap(0\times\mathbb{R}_{p}^{n}%
)=0$). Then $S$ admits a homogeneous free generating function
\[
W(x,x^{\prime})=\frac{1}{2}B^{-1}Ax^{2}-B^{-1}xx^{\prime}+\frac{1}{2}%
DB^{-1}x^{\prime2}%
\]
(where $B^{-1}Ax^{2}=B^{-1}Ax\cdot x$, etc.), and we have $(x_{S}%
,p_{S})=S(x,p)$ if and only if $p_{S}=\nabla_{x}W(x_{S},x)$ and $p=-\nabla
_{x}W(x_{S},x)$. Since $W$ is homogeneous of degree two in the $x,x^{\prime}$
variables, Euler's formula yields%
\begin{align*}
W(x_{S},x)  & =\frac{1}{2}(x_{S}\nabla_{x_{S}}W(x_{S},x)+x\nabla_{x}%
W(x_{S},x))\\
& =\frac{1}{2}(p_{S}x_{S}-px)
\end{align*}
hence formula (\ref{ph4}) can be rewritten as%
\begin{equation}
\varphi_{S}(\check{z})=\varphi(\check{z})+W(x_{S},x)\text{.}\label{ph6}%
\end{equation}

As we will see in Section \ref{tre} formula (\ref{ph6}) is a particular case
of a more general result describing the action of Hamiltonian flows on the
phase of a Lagrangian manifold.

\section{Hamiltonian Motions and Phase\label{tre}}

We are now going to investigate the action of Hamiltonian flows on the phase.
Let us first introduce some notations. Let $H=H(z,t)$ (\textquotedblleft the
Hamiltonian\textquotedblright) be a smooth real function defined on
$\mathbb{R}_{z}^{2n}\times\mathbb{R}_{t}$. We denote by $(f_{t,t^{\prime}}%
^{H})$ the \textit{time-dependent flow} it determines: for an initial point
$z^{\prime}$ set $z_{t}=f_{t,t^{\prime}}^{H}(z_{t^{\prime}})$; the function
$t\longmapsto z_{t}$ is the solution of Hamilton's equations%
\[
\dot{x}=\nabla_{p}H(z,t)\text{ \ , \ }\dot{p}=-\nabla_{x}H(z,t)(z,t)
\]
passing through $z_{t^{\prime}}^{\prime}$ at time $t^{\prime}$. Notice that
$f_{t,t^{\prime}}^{H}\circ f_{t^{\prime},t^{\prime\prime}}^{H}=f_{t,t^{\prime
\prime}}^{H}$. The \textit{suspended Hamiltonian flow} $(\tilde{f}_{t}^{H})$
is defined on the extended phase space $\mathbb{R}_{z}^{2n}\times
\mathbb{R}_{t}$; it is defined by
\[
\tilde{f}_{t}^{H}(z^{\prime},t^{\prime})=(f_{t+t^{\prime},t^{\prime}}%
^{H}(z^{\prime}),t+t^{\prime})
\]
and is the flow of the suspended Hamiltonian vector field
\[
\tilde{X}_{H}=(\nabla_{p}H,-\nabla_{x}H,1);
\]
we will also use the notation%
\[
f_{t}^{H}=f_{t,0}^{H}%
\]
and call, somewhat sloppily, the family of canonical transformations
$(f_{t}^{H})$ the\emph{ flow determined by} $H$ (it is not truly a flow when
$H$ is effectively time-dependent: $f_{t}^{H}f_{t^{\prime}}^{H}\neq
f_{t+t^{\prime}}^{H}$).

We will use the properties of the Poincar\'{e}--Cartan integral form. It is
the one-form $\alpha_{H}$ on $\mathbb{R}_{z}^{2n}\times\mathbb{R}_{t}$ defined
by
\[
\alpha_{H}=p\mathrm{d}x-H\mathrm{d}t\text{.}%
\]
Its interest comes from the following property which expresses the fact
$\alpha_{H}$ is a relative integral invariant (see \cite{AM,Marle}): the
contraction of the exterior derivative
\[
\mathrm{d}\alpha_{H}=\mathrm{d}p\wedge\mathrm{d}x-\mathrm{d}H\wedge\mathrm{d}t
\]
with the suspended Hamilton vector field $\tilde{X}_{H}$ is zero:
$i_{\tilde{X}_{H}}\mathrm{d}\alpha_{H}=0$. This means that for every vector
$\tilde{Y}(z,t)$ in $\mathbb{R}_{z}^{2n}\times\mathbb{R}_{t}$ originating at a
point $(z,t)$ we will have
\begin{equation}
\mathrm{d}\alpha_{H}(\tilde{X}_{H}(z,t),\tilde{Y}(z,t))=0\text{.}\label{cont}%
\end{equation}
This property has the following, for us very important, consequence: let
$\tilde{\gamma}:[0,1]\longrightarrow\mathbb{R}_{z}^{2n}\times\mathbb{R}_{t}$
be a smooth curve in extended phase space on which we let the suspended flow
$\tilde{f}_{t}^{H}$ act; as time varies, $\tilde{\gamma}$ will sweep out a
two-dimensional surface $\Sigma_{t}$ whose boundary $\partial\Sigma_{t}$
consists of $\tilde{\gamma}$, $\tilde{f}_{t}^{H}(\tilde{\gamma})$, and two
arcs of phase-space trajectory, $\tilde{\gamma}_{0}$ and $\tilde{\gamma}_{1}$:
$\tilde{\gamma}_{0}$ is the trajectory of the origin $\tilde{\gamma}(0)$ of
$\tilde{\gamma}$, and $\tilde{\gamma}_{1}$ that of its endpoint $\tilde
{\gamma}(1)$. It turns out that we will have%
\begin{equation}
\int_{\partial\Sigma_{t}}\alpha_{H}=0\text{.}\label{fund}%
\end{equation}
Here is a sketch of the proof: using the multi-dimensional Stokes formula we
have%
\[
\int_{\partial\Sigma_{t}}\alpha_{H}=\int_{\Sigma_{t}}\mathrm{d}\alpha
_{H}\text{.}%
\]
Since the surface $\Sigma_{t}$ consists of flow lines of $\tilde{X}_{H}$ each
pair $(\tilde{X},\tilde{Y})$ of tangent vectors at a point $(z,t)$ can be
written as a linear combination of two independent vectors, and one of these
vectors can be chosen as $\tilde{X}_{H}$. It follows that $\mathrm{d}%
\alpha_{H}(\tilde{X},\tilde{Y})$ is a sum of terms of the type $\mathrm{d}%
\alpha_{H}(\tilde{X}_{H},\tilde{Y})$, which are equal to zero in view of
(\ref{cont}). We thus have $\int_{\Sigma_{t}}\mathrm{d}\alpha_{H}=0$, whence
(\ref{fund}).

From now on we will use the definite integral notation
\[
\int_{z_{t^{\prime}}}^{z_{t}}\alpha_{H}=\int_{z_{t^{\prime}}}^{z_{t}%
}p\mathrm{d}x-H\mathrm{d}t
\]
for the integral of the Poincar\'{e}--Cartan form along the phase space
trajectory $s\longmapsto f_{s,t^{\prime}}^{H}(z_{t^{\prime}})$ joining
$z_{t^{\prime}}$ to $z_{t}=f_{t,t^{\prime}}^{H}(z_{t^{\prime}})$. Let
$\mathbb{V}$ be a Lagrangian manifold and $\mathbb{V}_{t}=f_{t}^{H}%
(\mathbb{V)}$. Notice that $\mathbb{V}_{t}=f_{t,t^{\prime}}^{H}(\mathbb{V}%
_{t^{\prime}})$. Since Hamiltonian flows consist of symplectomorphisms each
$\mathbb{V}_{t}$ is a Lagrangian manifold, and the function $\varphi
_{t}:\mathbb{\check{V}}_{t}\longrightarrow\mathbb{R}$ defined by%
\[
\varphi_{t}(\check{z}_{t})=\int_{\gamma(\bar{z}_{t},z_{t})}p\mathrm{d}x
\]
($\check{z}_{t}$ being the homotopy class in $\mathbb{V}_{t}$ of a path
$\gamma(\bar{z}_{t},z_{t})$) obviously is a phase when $\bar{z}_{t}=f_{t}%
^{H}(\bar{z}_{0})$ is chosen as base point in $\mathbb{V}_{t}$. The following
result relates $\varphi_{t}$ to the phase $\varphi=\varphi_{0}$ of
$\mathbb{V}=\mathbb{V}_{0}$:

\begin{lemma}
\label{lemab}Let $\check{z}=\check{z}_{0}$ be a point in $\mathbb{V}$ and
$\check{z}_{t}$ its image in $\mathbb{V}_{t}$ (i.e. $\check{z}_{t}$ is the
homotopy class in $\mathbb{V}_{t}$ of the image by $f_{t}^{H}$ of a path
representing $\check{z}$).We have%
\begin{equation}
\varphi_{t}(\check{z}_{t})-\varphi(\check{z})=\int_{\bar{z}_{0}}^{\bar{z}_{t}%
}\alpha_{H}-\int_{z_{0}}^{z_{t}}\alpha_{H}\text{.}\label{fif}%
\end{equation}

\end{lemma}

\begin{proof}
Let $\Sigma_{t}$ be the closed curve
\[
\Sigma_{t}=[\bar{z}_{0},\bar{z}_{t}]+\gamma(\bar{z}_{t},z_{t})-[z_{0}%
,z_{t}]-\gamma(\bar{z}_{0},z_{0})
\]
where $[\bar{z}_{0},\bar{z}_{t}]$ (resp. $[z_{0},z_{t}]$)\ is the Hamiltonian
trajectory joining $\bar{z}_{0}$ to $\bar{z}_{t}$ (resp. $z_{0}$ to $z_{t}$).
In view of the consequence (\ref{fund}) of the relative invariance property of
the Poincar\'{e}--Cartan form $\alpha_{H}$ we have%
\begin{equation}
\int_{\Sigma_{t}}\alpha_{H}=0\text{.}\label{sigmat}%
\end{equation}
Since $\mathrm{d}t=0$ along both $\gamma(\bar{z}_{t},z_{t})$ and $\gamma
(\bar{z}_{0},z_{0})$ we have%
\[
\int_{\gamma(\bar{z}_{t},z_{t})}\alpha_{H}=\int_{\gamma(\bar{z}_{t},z_{t}%
)}p\mathrm{d}x\text{ \ \ , \ }\int_{\gamma(\bar{z}_{0},z_{0})}\alpha_{H}%
=\int_{\gamma(\bar{z}_{0},z_{0})}p\mathrm{d}x
\]
and hence (\ref{sigmat}) is equivalent to
\[
\int_{\gamma(\bar{z}_{0},z_{0})}p\mathrm{d}x+\int_{z_{0}}^{z_{t}}\alpha
_{H}-\int_{\gamma(\bar{z}_{t},z_{t})}p\mathrm{d}x+\int_{\bar{z}_{0}}^{\bar
{z}_{t}}\alpha_{H}=0
\]
that is to (\ref{fif}).
\end{proof}

Lemma \ref{lemab} has the following fundamental consequence for the phase of
$\mathbb{V}_{t}=f_{t}^{H}(\mathbb{V)}$:

\begin{proposition}
\label{phapha}Set $z_{0}=z$ and $\check{z}=\check{z}_{0}$. The function
$\varphi(\cdot,t):\mathbb{\check{V}}\longrightarrow\mathbb{R}$ defined by%
\begin{equation}
\varphi(\check{z},t)=\varphi(\check{z})+\int_{z,0}^{z_{t},t}\alpha_{H}\text{
\ , \ \ }z_{t}=f_{t}^{H}(z)\label{ph7}%
\end{equation}
is a phase of $\mathbb{V}_{t}$: for fixed $t$ we have%
\begin{equation}
\mathrm{d}\varphi(\check{z},t)=p_{t}\mathrm{d}x_{t}\text{ \ \ \textrm{if}
\ }\pi_{t}(\check{z})=z_{t}=(x_{t},p_{t})\label{pha}%
\end{equation}
that is, equivalently,%
\begin{equation}
\mathrm{d}\varphi(\check{z},t)=p_{t}\mathrm{d}x_{t}\text{ \ \ \textrm{if}
\ }\pi(\check{z})=z=(x,p)\text{.}\label{phb}%
\end{equation}

\end{proposition}

\begin{proof}
In view of Lemma \ref{lemab} the function $\varphi(\cdot,t)$ satisfies
\begin{equation}
\varphi(\check{z},t)=\varphi_{t}(\check{z}_{t})+\int_{\bar{z}_{0}}^{\bar
{z}_{t}}\alpha_{H}\label{ph5}%
\end{equation}
where $\check{z}_{t}\in\mathbb{\check{V}}_{t}$ is the image of $\check{z}$ by
$f_{t}^{H}$. It follows that for fixed $t$ we have%
\[
\mathrm{d}\varphi(\check{z},t)=p_{t}\mathrm{d}x_{t}\text{ \ \textrm{if} \ }%
\pi_{t}(\check{z}_{t})=(x_{t},p_{t})
\]
where $\pi_{t}:\mathbb{\check{V}}_{t}\longrightarrow\mathbb{V}_{t}$ is the
projection $\check{z}_{t}\longmapsto z_{t}$.
\end{proof}

We will call the function $\varphi(\cdot,t):\mathbb{\check{V}}\longrightarrow
\mathbb{R}$ the phase of $\mathbb{V}_{t}$; observe that it is defined, not on
$\mathbb{\check{V}}_{t}$, but on $\mathbb{\check{V}}$ itself, viewed as a
\textquotedblleft master universal covering manifold\textquotedblright.

The following particular case relates the Hamiltonian phase to Proposition
\ref{un} on changes of symplectic frames:

\begin{corollary}
Let $H$ be a Hamiltonian which is quadratic and homogeneous in the position
and momentum variables; its flow thus consists of symplectic matrices
$S_{t}^{H}$. The Hamiltonian phase of $S_{t}^{H}(\mathbb{V)}$ is
\begin{equation}
\varphi(\check{z},t)=\varphi(\check{z})+\tfrac{1}{2}(p_{t}x_{t}-px)\text{.}%
\label{stv}%
\end{equation}

\end{corollary}

\begin{proof}
Since $H$ is quadratic we have, using successively Euler's formula and
Hamilton's equations,%
\begin{align*}
H(z_{t},t)  & =\tfrac{1}{2}(x_{t}\nabla_{x}H(z_{t},t)+p_{t}\nabla_{p}%
H(z_{t},t))\\
& =\tfrac{1}{2}(-x_{t}\dot{p}_{t}+p_{t}\dot{x}_{t})
\end{align*}
and hence%
\begin{align*}
\int_{z_{0}}^{z_{t}}\alpha_{H}  & =\tfrac{1}{2}\int_{0}^{t}(p_{s}\dot{x}%
_{s}+x_{s}\dot{p}_{s})\mathrm{d}s\\
& =\tfrac{1}{2}(p_{t}x_{t}-px)
\end{align*}
whence (\ref{stv}) in view of (\ref{ph7}).
\end{proof}

Another interesting particular case of Proposition \ref{phapha} occurs when
the Lagrangian manifold $\mathbb{V}$ is invariant under the flow: $f_{t}%
^{H}(\mathbb{V)}=\mathbb{V}$. (This situation typically occurs when one has a
completely integrable system and $\mathbb{V}$ is an associated Lagrangian torus.)

\begin{corollary}
\label{coro}Let $H$ be a time-independent Hamiltonian, $(f_{t}^{H})$ its flow,
and assume that $\mathbb{V}$ is invariant under $(f_{t}^{H})$ (that is
$f_{t}^{H}(\mathbb{V)}=\mathbb{V}$ \ for all $t$). If $\check{z}$ is the
homotopy class in $\mathbb{V}$ of a a path $\gamma(\bar{z}_{0},z)$ and
$\gamma(z,z_{t})$ is the piece of Hamiltonian trajectory joining $z$ to
$z_{t}$ then
\begin{equation}
\varphi(\check{z},t)=\varphi(\check{z}_{t})-Et\label{phzt}%
\end{equation}
where $E$ is the (constant) value of $H$ on $\mathbb{V}$ and $\check{z}_{t}$
the homotopy class of the path $\gamma(\bar{z}_{0},z)+\gamma(z,z_{t})$ in
$\mathbb{V}$.
\end{corollary}

\begin{proof}
The trajectory $s\longmapsto$ $z_{s}=f_{s}^{H}(z)$ is a path $\gamma(z,z_{t})$
in $\mathbb{V}$ joining $z$ to $z_{t}$ hence%
\[
\int_{\gamma(\bar{z}_{0},z)}p\mathrm{d}x+\int_{z}^{z_{t}}\alpha_{H}%
=\int_{\gamma(\bar{z}_{0},z_{t})}p\mathrm{d}x-Et
\]
where $\gamma(\bar{z}_{0},z_{t})=\gamma(\bar{z}_{0},z)+\gamma(z,z_{t})$. The
result follows since the first integral in the right-hand side of this
equality is by definition $\varphi(\check{z}_{t})$.
\end{proof}

Proposition \ref{phapha} allows us also to link the notion of phase of a
Lagrangian manifold to the standard Hamilton--Jacobi theory.

\begin{proposition}
Let $z\in\mathbb{V}$ have a neighborhood $\mathbb{U}$ in $\mathbb{V}$
projecting diffeomorphically on $\mathbb{R}_{x}^{n}$. $(\ast)$ There exists
$\varepsilon>0$ such that the local expression $\Phi=\Phi(x,t)$ of the phase
$\varphi$ is defined for $|t|<\varepsilon$ and $(\ast\ast)$ $\Phi$ satisfies
the Hamilton--Jacobi equation%
\[
\frac{\partial\Phi}{\partial t}+H(x,\nabla_{x}\Phi)=0
\]
for $|t|<\varepsilon$.
\end{proposition}

\begin{proof}
The first part $(\ast)$ is an immediate consequence of Proposition (\ref{fi})
(the existence of $\varepsilon$ follows from the fact that the caustic is a
closed subset of $\mathbb{V}$). To prove $(\ast\ast)$ we observe that%
\[
\Phi(x,t)=\Phi(x^{\prime},0)+\int_{z^{\prime},0}^{z,t}p\mathrm{d}%
x-H\mathrm{d}t
\]
in view of formula (\ref{ph7}) in Proposition \ref{phapha}; now we can
parametrize the arc joining $z^{\prime},0$ to $z,t$ by $x$ and $t$ hence%
\[
\Phi(x,t)=\Phi(x^{\prime},0)+\int_{x^{\prime},0}^{x,t}p\mathrm{d}%
x-H\mathrm{d}t
\]
which is precisely the solution of Hamilton--Jacobi's equation with initial
datum $\Phi$ at time $t=0$ (cf. formula (\ref{solution})).
\end{proof}

\section{Phase and Heisenberg--Weyl Operators\label{qua}}

Let $T(z_{a})$ be the phase space translation $z\longmapsto z+z_{a}$. That
operator can be viewed as the time-one map of the flow determined by the
\textquotedblleft translation Hamiltonian $H^{a}=\sigma(z,z_{a})$: that flow
consists of the mappings $f_{t}^{a}(z)=z+tz_{a}$, and thus $T(z_{a})=f_{1}%
^{a}$. In Proposition \ref{hewey} below this is taken into account, that is,
the phases of the translated Lagrangian manifolds will be calculated using
formula (\ref{ph7}) of Proposition \ref{phapha} above.

\begin{proposition}
\label{hewey}Let $T(z_{a})$ be the translation with vector $z_{a}=(x_{a}%
,p_{a})$ and $T(z_{b})$ that with vector $z_{b}=(x_{b},p_{b})$. $(\ast)$ The
phase $\varphi_{a}$ of $T(z_{a})\mathbb{V}$ is given by%
\begin{equation}
\varphi_{a}(\check{z})=\varphi(\check{z})+\tfrac{1}{2}p_{a}x_{a}+p_{a}%
x_{0}\text{ \ \textrm{if} \ }\pi(\check{z})=(x,p)\text{.}\label{eg1}%
\end{equation}
$(\ast\ast)$ Let $\varphi_{a,b}$ be the phase of $T(z_{a})(T(z_{b}%
)\mathbb{V)}$ and $\varphi_{a+b}$ that of $T(z_{a}+z_{b})\mathbb{V}$; we have
\begin{equation}
\varphi_{a,b}(\check{z})-\varphi_{a+b}(\check{z})=-\tfrac{1}{2}\sigma
(z_{a},z_{b})\label{eg2}%
\end{equation}
and hence%
\begin{equation}
\varphi_{a,b}(\check{z})-\varphi_{b,a}(\check{z})=\sigma(z_{a},z_{b}%
)\text{.}\label{eg3}%
\end{equation}

\end{proposition}

\begin{proof}
We have, in view of (\ref{ph5}),
\begin{align*}
\varphi_{a}(\check{z})  & =\varphi(\check{z}_{0})+\int_{0}^{1}(p_{0}%
+tp_{a})x_{a}\mathrm{d}t-\int_{0}^{1}\sigma(z_{0}+tz_{a},z_{a})\mathrm{d}t\\
& =\varphi(\check{z}_{0})+p_{0}x_{a}+\tfrac{1}{2}p_{a}x_{a}-(p_{0}x_{a}%
-p_{a}x_{0})\\
& =\varphi(\check{z}_{0})+\tfrac{1}{2}p_{a}x_{a}+p_{a}x_{0}%
\end{align*}
whence (\ref{eg1}). Formulae (\ref{eg2}) follows from (\ref{eg1}) since we
have
\[
\varphi_{a,b}(\check{z})=(\tfrac{1}{2}p_{b}x_{b}+p_{b}x_{0})+(\tfrac{1}%
{2}p_{a}x_{a}+p_{a}(x_{b}+x_{0}))
\]
and%
\[
\varphi_{a+b}(\check{z})=\tfrac{1}{2}(p_{a}+p_{b})(x_{a}+x_{b})+(p_{a}%
+p_{b})x_{0}\text{.}%
\]
Formula (\ref{eg3}) follows from formula (\ref{eg2}).
\end{proof}

\begin{remark}
The phases of $T(z_{a})(T(z_{b})\mathbb{V)}$ and $T(z_{a}+z_{b})\mathbb{V}$
are different, even though these manifolds are\textit{\ the same}! In fact,
formula (\ref{eg2}) shows that the difference between the phases of
$T(z_{a}+z_{b})\mathbb{V}$ and $T(z_{a})(T(z_{b})\mathbb{V)}$ is just (up to
the sign) the area of the phase space triangle spanned by the vectors
$z_{a},z_{b}$ (see the discussion and Fig.3, p.211 in Littlejohn
\cite{Littlejohn}).
\end{remark}

We also have the following \textquotedblleft symplectic
covariance\textquotedblright\ result:

\begin{proposition}
\label{syco}The Hamiltonian phases of the identical Lagrangian manifolds
$S_{t}^{H}(T(z_{a})\mathbb{V)}$ and $T(S_{t}^{H}(z_{a}))S_{t}^{H}\mathbb{V}$
are equal.
\end{proposition}

\begin{proof}
The phase of $T(z_{a})\mathbb{V}$ is
\[
\varphi_{a}(\check{z})=\varphi(\check{z}_{0})+\tfrac{1}{2}p_{a}x_{a}%
+p_{a}x_{0}%
\]
hence that of $S_{t}^{H}(T(z_{a})\mathbb{V)}$ is (using (\ref{stv}) and the
linearity of $S_{t}^{H}$):%
\begin{multline*}
A(t)=\varphi(\check{z}_{0})+\tfrac{1}{2}p_{a}x_{a}+p_{a}x_{0}+\\
\tfrac{1}{2}(p_{0,t}+p_{a,t})(x_{0,t}+x_{a,t})-\tfrac{1}{2}(p_{0}+p_{a}%
)(x_{0}+x_{a}))
\end{multline*}
where $z_{0,t}=S_{t}^{H}z_{0}$, $z_{a,t}=S_{t}^{H}z_{a}$. Similarly, the
Hamiltonian phase of $S_{t}^{H}\mathbb{V}$ is%
\[
\varphi(\check{z},t)=\varphi(\check{z}_{0})+\tfrac{1}{2}(p_{0,t}x_{0,t}%
-p_{0}x_{0})
\]
hence that of $T(S_{t}^{H}(z_{a}))S_{t}^{H}\mathbb{V}$ is%
\[
B(t)=\varphi(\check{z}_{0})+\tfrac{1}{2}(p_{0,t}x_{0,t}-p_{0}x_{0})+\tfrac
{1}{2}p_{a,t}x_{a,t}+p_{a,t}x_{0,t}%
\]
and thus%
\begin{align*}
A(t)-B(t)  & =\tfrac{1}{2}(p_{a}x_{0}-p_{0}x_{a})-\tfrac{1}{2}(p_{a,t}%
x_{0,t}-p_{0,t}x_{a,t})\\
& =\tfrac{1}{2}(\sigma(z_{a},z_{0})-\sigma(z_{a,t},z_{0,t}))\\
& =\tfrac{1}{2}(\sigma(z_{a},z_{0})-\sigma(S_{t}^{H}z_{a},S_{t}^{H}%
z_{0}))\text{.}%
\end{align*}
Since $S_{t}^{H}\in\operatorname*{Sp}(n)$ we have $\sigma(S_{t}^{H}z_{a}%
,S_{t}^{H}z_{0})=\sigma(z_{a},z_{0})$ and hence $A(t)=B(t)$, proving the Proposition.
\end{proof}

Let us extend the results above to the case of a Hamiltonian flow
\textquotedblleft displacing\textquotedblright\ points in the direction of the
field of tangents to a smooth curve in phase space. To such a curve
$t\longmapsto\gamma(t)=(x^{\gamma}(t),p^{\gamma}(t))$ we associate the
time-dependent Hamiltonian $H^{\gamma}$ defined by%
\[
H^{\gamma}(z,t)=\sigma(z,\dot{\gamma}(t))=p\dot{x}^{\gamma}(t)-x\dot
{p}^{\gamma}(t)\text{.}%
\]
The solutions of the associated Hamilton equations $\dot{x}(t)=\dot{x}%
^{\gamma}(t)$,\ $\dot{p}(t)=\dot{p}^{\gamma}(t)$ are given by%
\[
z_{t}=z(0)+\gamma(t)-\gamma(0)
\]
hence the flow $(f_{t,t^{\prime}}^{\gamma})$ propagates points along curves
which are translations of $\gamma$. We will therefore call $H^{\gamma}$ the
displacement Hamiltonian along $\gamma$. Set $f_{t}^{\gamma}=f_{t,0}^{\gamma}$
and let $\mathbb{V}$ be a Lagrangian manifold with phase $\varphi$.

\begin{proposition}
The phase of the displaced Lagrangian manifold $\mathbb{V}^{\gamma(t)}%
=f_{t,0}^{\gamma}(\mathbb{V})$ is%
\begin{equation}
\varphi_{\gamma}(\check{z},t)=\varphi(\check{z})+\tfrac{1}{2}(p_{t}x_{t}%
-p_{0}x_{0})-\tfrac{1}{2}\int_{\gamma}p\mathrm{d}x-x\mathrm{d}p\text{.}%
\label{phg}%
\end{equation}
If, in particular, $\gamma$ is a loop then%
\begin{equation}
\varphi_{\gamma}(\check{z},t)=\varphi(\check{z})-\tfrac{1}{2}\int_{\gamma
}p\mathrm{d}x-x\mathrm{d}p\text{.}\label{phh}%
\end{equation}

\end{proposition}

\begin{proof}
In view of formula (\ref{ph7}) the phase of $\mathbb{V}^{\gamma(t)}$ is
\[
\varphi(\check{z},t)=\varphi(\check{z})+\int_{0}^{t}(p(s)\dot{x}%
(s)-\sigma(z(s),\dot{\gamma}(s)))\mathrm{d}s\text{;}%
\]
we have $\sigma(z(s),\dot{\gamma}(s))=\sigma(z(s),\dot{z}(s))$ and hence
\[
p(s)\dot{x}(s)-\sigma(z(s),\dot{\gamma}(s))=\dot{p}(s)x(s)\text{.}%
\]
Noting that%
\begin{align*}
\dot{p}(s)x(s)  & =\tfrac{1}{2}(p(s)\dot{x}(s)+\dot{p}(s)x(s)-(p(s)\dot
{x}(s)-\dot{p}(s)x(s)))\\
& =\tfrac{1}{2}\tfrac{\mathrm{d}}{\mathrm{d}t}(p(s)x(s))-\tfrac{1}{2}%
\sigma(z(s),\dot{z}(s))
\end{align*}
and integrating we get formula (\ref{phg}).
\end{proof}

The result above can actually be recovered from Proposition (\ref{hewey}) by
using infinitesimal translations: segmenting the trajectory $t\longmapsto
z(t)$ into straight sections $[z,z_{1}]$, $[z_{1},z_{2}]$,... where
$z_{k}=z(k\Delta t)$ $(\Delta t=t/N)$, one finds that the limit of the product
of these operators is precisely
\[
\lim_{N\rightarrow\infty}T(z_{N}-z_{N-1})\cdot\cdot\cdot T(z_{2}-z_{1}%
)T(z_{1}-z)=T^{\gamma(t)}\text{.}%
\]
(This observation thus a posteriori justifies formula (3.27), p.212, in
Littlejohn \cite{Littlejohn}.)

Both formulas (\ref{eg2}), (\ref{eg3}) in Proposition \ref{hewey} are strongly
reminiscent of the commutation formulas in the quantum-mechanical
Heisenberg--Weyl group; however there is nothing quantum mechanical involved
in our constructions! Let us discuss this point in some detail. Recall (see
for instance Littlejohn \cite{Littlejohn}) that the basic idea of the
Heisenberg--Weyl operators is that they move wave functions around in phase
space. This is done as follows: for a given quantum state $\left\vert
\Psi\right\rangle $ the position and momentum expectation values are
$\left\langle x\right\rangle $ and $\left\langle p\right\rangle $; this can be
written collectively as $\left\langle z\right\rangle =\left\langle
\Psi\right\vert z\left\vert \Psi\right\rangle $. Heisenberg--Weyl operators
$\hat{T}(z_{a})$ are parameterized by points $z_{a}$ in phase space, and have
the property that if $\left\vert \Psi\right\rangle $ has the expectation value
$\left\langle z\right\rangle $ then $\hat{T}(z_{a})\left\vert \Psi
\right\rangle $ should have the expectation value $\left\langle z\right\rangle
+z_{a}$; this requires that
\[
\hat{T}(z_{a})^{\ast}\hat{z}\hat{T}(z_{a})=\hat{z}+z_{a}%
\]
where $\hat{z}=(x,-i\hbar\nabla_{x})$ is the quantum operator associated with
$z$. One shows that this implies that $\hat{T}(z_{a})$ must be the operator
\[
\hat{T}(z_{a})=\exp(\tfrac{i}{\hbar}\sigma(z_{a},\hat{z}))
\]
whose action on wave functions in the $x$-representation is given by%
\[
\hat{T}(z_{a})\Psi(x)=\exp(\tfrac{i}{\hbar}(p_{a}x-\tfrac{1}{2}p_{a}x_{a}%
)\Psi(x-x_{a})\text{.}%
\]

Let us interpret Propositions \ref{hewey} and \ref{syco} in terms of the wave
functions
\[
\Psi(\check{z})=\exp(\tfrac{i}{\hbar}\varphi(\check{z}))\sqrt{\rho}(\check{z})
\]
on $\mathbb{\check{V}}$ introduced in our previous work \cite{paselect,ICP};
$\rho$ is here a de Rham form on $\mathbb{\check{V}}$. Such a wave-function is
defined on $\mathbb{V}$, i.e.
\[
\Psi(\gamma\check{z})=\Psi(\check{z})\text{ \ for all \ }\gamma\in\pi
_{1}(\mathbb{V})
\]
if and only if $\mathbb{V}$ satisfies the \textit{EBK} condition
\[
\frac{1}{2\pi\hbar}\oint\nolimits_{\gamma}p\mathrm{d}x-\frac{1}{4}%
m(\gamma)\text{ \ is an integer.}%
\]
($\gamma$ an arbitrary loop on $\mathbb{V}$, $m(\gamma)$ its Maslov index). We
define the action of the Hamiltonian flow $(f_{t}^{H})$ on $\Psi$ by%
\[
f_{t}^{H}\Psi(\check{z})=\exp(\tfrac{i}{\hbar}\varphi(\check{z},t))\sqrt{\rho
}(\check{z},t)
\]
where $\varphi(\check{z},t)$ is the Hamiltonian phase and $\sqrt{\rho}%
(\check{z},t)=\sqrt{\rho}(f_{t}^{H}(\check{z}),t)$. If we now choose for $H$
the translation Hamiltonian $H^{a}(z)=\sigma(z,z_{a})$ then in view of
(\ref{eg1}) the action of $T(z_{a})=f_{1}^{H^{a}}$ on $\Psi$ is
\[
T(z_{a})\Psi(\check{z})=\exp[\tfrac{i}{\hbar}(\varphi(\check{z})+\tfrac{1}%
{2}p_{a}x_{a})]\sqrt{\rho}(T(z_{a})\check{z})\text{.}%
\]
This action of translation operators satisfies, in view of (\ref{eg2}),
(\ref{eg3}) the commutation relations%
\[
T(z_{b})T(z_{a})\Psi(\check{z})=T(z_{a})T(z_{b})\Psi(\check{z})\text{.}%
\]

\end{document}